# Polymetric Rhythmic Feel for a Cognitive Drum Computer

Oliver Weede

Hochschule für Kunst, Design und Populäre Musik (hKDM)
Freiburg, Germany
oliver.weede@hkdm.de

**Abstract**

This paper addresses a question about music cognition: how do we derive polymetric structures. A preference rule system is presented which is implemented into a drum computer. The preference rule system allows inferring local polymetric structures, like two-over-three and three-over-two. By analyzing the micro-timing of West African percussion music a timing pattern consisting of six pulses was discovered. It integrates binary and ternary rhythmic feels. The presented drum computer integrates the discovered superimposed polymetric swing (timing and velocity) appropriate to the rhythmic sequence the user inputs. For binary sequences, the amount of binary swing is increased and for ternary sequences, the ternary swing is increased.

## 1 Introduction

Traditionally, it is the task of a human performer to add artistic interpretation in order to bring the composition to life and to allow it to sound. The performer adds "feeling", expression, articulation, phrasing or swing. If we describe the composer as a sender, the musical performer as a transfer element and the recipient as a receiver, a lack of expressivity arises when the performer is missing. A composition based on simple common music notation played exactly as written sounds rigid and mechanical. An aim of the current work is to model artistic expression and cognitive processes of a human musical performer to obtain a vivid interpretation by the computer. The second aim is to gain a deeper knowledge of polyrhythmic music and timing structures. Up to date, digital audio workstations allow to add binary swing and to randomize velocities and note onsets. In some systems, there is the possibility to copy a swing-pattern from an example to other sequences. However, the solution is not based on a musical understanding. The result of the "humanization" functions sounds livelier, but a recipient quickly identifies the arbitrariness of variations or an overlaid rigid structure. Instead of learning by direct imitation, like it is e.g. described by Tidemann et al. [Tidem08], the approach presented here is based on a model of the human perception of music and on results of musical performance analysis. Rhythmic sequences, entered by the user, are analyzed to identify the metric structure and rhythmic phrases. These rhythms are then played with an appropriate feel: micro timing and velocity.

## 1.1 Music Analysis

The approach for rhythmic analysis is based on an adapted model of the preference rule system for the cognition of basic musical structures of David Temperley [Temperl01]. In addition to Temperley's model, in which one global meter is inferred, in this approach, local polymetric structures are identified for each rhythmic instrument of a musical ensemble. These local metric structures can be the result of temporary ternary sequences in a global binary meter (three-over-two) or of binary sequences in a ternary meter (two-over-three).

## 1.2 Micro Timing

Variations in timing and velocity are a central quality of rhythmical musical performance. Micro timing describes timing variations at a constant tempo. While the beats of the metric structure are isochronous, the beat subdivisions are not. In jazz, micro timing variations and the corresponding rhythmic feel is referred to as swing. Consecutive eighth-notes are performed as long-short patterns. Butterfield [Butterf11] states that swing also includes varying degrees of velocity to cause vital drive, motional energy and the sense of forward propulsion by „pulling against the pulse". Pulses have been regarded as isochronous time spans to represent a time grid at the lowest (or subdivision) level of meter. However, Polak states that non-isochronous subdivisions of the beat in terms of stable timing patterns are fundamental to a metric system [Polak10]. Thus, pulses can be non-isochronous. London [London04] states that lengthenings and shortenings are not deviations from the norm, they are the norm. In this sense, micro timing is no derivation, but rather an inflected non-isochronous pulse of the meter.

While stable timing patterns of two or three pulses are well known, surprisingly not much is written about timing patterns of length six that arise by superimposing both patterns. In this paper, the concept of a superimposed polymetric swing is introduced to describe a polymetric timing pattern at the level of pulses.

## 2    Micro Timing

### 2.1    Timing patterns as rhymic feels

There are at least two categories of micro timing: binary and ternary swing. In *binary swing*, the cyclic repeating timing pattern consists of two pulses. The onset of each uneven pulse (offbeat) is inflected while the even pulses (downbeats) are played exactly on an isochronous grid. In other words, the first note of a binary division is varied in length. In an eighth swing notated in a 4/4 meter the inflected onset is the offbeat eighth-note. Most commonly it is delayed, resulting in an "iambic swing". In this case, the offbeat serves as an anacrusis for the strong beat; the forward-propelling directionality of the delayed note links it together with the following downbeat and increases the motional energy; the anacrusis is oriented to a new beginning, it is directed toward a future event rather than completing a phrase [Butterf11]. Dependent on the inflection, binary swing leads to a beat subdivision of a long and a short pulse (LS) or - less common - to a short pulse on the downbeat followed by a long offbeat (trochee meter). In Fig. 1 (a) the "iambic swing" is shown; each uneven pulse is delayed.

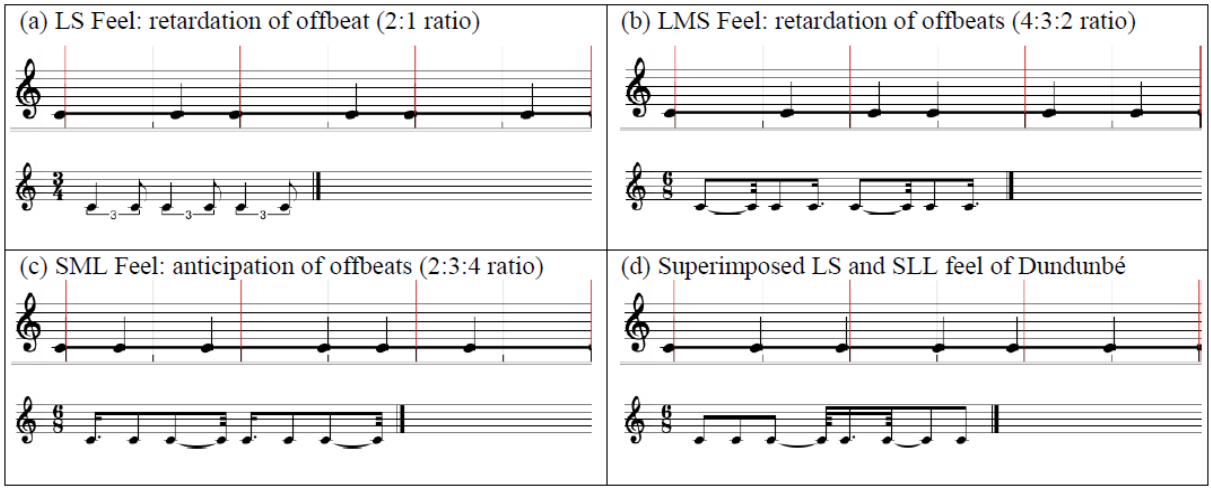

Fig. 1: Example timing patterns. Pulse lengths (a) Long-Short (b) Long-Mid-Short (c) Short-Mid-Long (d) Mid-Mid-Long-Short-Long-Mid. The score is quantized to minimum units of sixty-fourth notes.

*Ternary swing* [Polak10] consists of a timing pattern with the length of three pulses. Like in binary swing, the length of the downbeat is inflected, but compared with binary swing there are two offbeats.

If the first offbeat pulse is pushed backward (*ternary swing with offbeat retardation*), the subdivisions are a long pulse, followed by a medium pulse (not changed in length) and a short pulse when the second offbeat is pushed backward with the same amount as the first offbeat (c.f. LMS feel in Fig. 1 (b)). Like the binary swing with offbeat eighth retardation, this ternary swing leads to an *anacrusis*: both offbeats are linked to the following downbeat (anapest). If the second offbeat is not delayed, the structure is Long-Short-Medium (LSM). In between LMS and LSM, there is the LSS structure. According to Polak's research [Polak10], stable LMS and LSS feels are important timing patterns in many rhythms of Mande music from Mali. The analysis performed in this work shows also that the LSM version is played in Mande music. The principal component of this class of ternary swing is that the downbeat is lengthened. So, we can call this swing *LFF feel* (Long-Flexible-Flexible).

The second class of ternary swing is one with *offbeat retardation*. If the first offbeat is played earlier, the resulting pulse lengths are SML (c.f. Fig. 1 (c)) or, if the second offbeat is delayed, SLL (c.f. [Polak10]) or SLM. In poetry, an SML pattern is called dactyl meter. Again, all three variations can be observed. The crux of this feel-class is that the downbeat is shortened. So, we can call this swing *SFF feel* (Short-Flexible-Flexible).

Despite there is the possibility to write down the score in a binary or ternary meter by using triplets or by increasing the resolution of notes, there is a structural difference between binary and ternary swing. The length of the cyclic repeating time pattern for binary swing is two, whereas the timing pattern of ternary swing has the length of three. In general, it consists of three different categories of pulse lengths instead of two. Superimposing both kinds of swing leads to a timing pattern of length six, where all five pulses followed by the first pulse are inflected (c.f. Fig. 1 (d)). The hypothesis is that in African music or music with African heritage, these pure binary or ternary versions are special cases of a *superimposed polymetric swing*. The norm is a mixture of the ternary and the binary feel, a timing pattern of six pulses.

## 2.2  Timing Measurements

To verify the hypothesis that micro timing might be a superimposed binary/ternary swing, I focused on analyzing polymetric West African percussion music of the Malinke from Guinea and southern regions of Mali, particularly recorded Djembé-performances of the master drummers Billy Nankouma Konaté and his father Famoudou Konaté[1]. 900 pulses have been measured. The empirically observed timing structures are determined by measuring time spans between the inter-onset-intervals of consecutive drum strokes. The measurements of each bar are normalized to a reference duration of one bar to compare timing patterns at different tempos. One rhythm of the LFF feel class was chosen, namely *Djaa*, and one rhythm with SFF feel class, namely *Dundunbé*. In both rhythms, binary and ternary sequences were studied to analyze whether the parts can be distinguished in their micro timing.[2]

## 2.3  Timing Model

A parametric model is introduced to verify the hypothesis that ternary and binary timing structures are superimposed. The optimal values for the parameters are computed by minimizing the sum of squared errors of the model. The prediction of the model is compared with the confidence intervals of the measurements.

---

[1] Famoudou Konaté is the first solo player of *Les Ballets Africains de la République de Guinée*. The recordings were taken with a zoom H4 handy recorder in Guinea and at workshops in Germany.

[2] The ternary part was a sequence of Tone, Slap, Slap (TSS); the binary part was TTSSSS.

An isochronous subdivision divides the beat into a downbeat and an offbeat with a ratio of 1:1. If a beat is regarded as a reference, we obtain the normed pulse lengths (0.5, 0.5), where the unit of both time spans is the "normed beat".[3] If the beat-offbeat-ratio is 2:1, the normed pulse lengths are (2/3, 1/3). In this case, a binary notated sequence could be regarded as ternary where the first offbeat of the ternary rhythm is always omitted (c.f. score in Fig. 1 (b)). However, the swing is binary. The delay of the offbeat is 1/6 normed beats. The downbeat has a length of 1/2+1/6 normed beats and the upbeat 1/2-1/6. The derivation from the isochronous pulse lengths are (+1/6, -1/6). To create a mathematical model, a binary swing factor $\theta_1$ is introduced which leads to the common 2:1 beat-offbeat-ratio for $\theta_1=1$. A factor of $\theta_1=0$ leads to the "straight" 1:1 ratio. The normed pulse lengths are then

$$\left(\frac{1}{2}+\frac{1}{6}\theta_1, \frac{1}{2}-\frac{1}{6}\theta_1\right)^T. \qquad (1)$$

First, just a factor $\theta_2$ was introduced to model the first offbeat of a ternary meter to obtain LFF versus SFF feel. The normed pulse lengths are

$$\left(\frac{1}{3}+\frac{1}{6}\theta_2, \frac{1}{3}, \frac{1}{3}-\frac{1}{6}\theta_2\right)^T. \qquad (2)$$

The derivation from the isochronous pulse lengths are $(1/6\theta_2, 0, -1/6\theta_2)$. For $\theta_2 > 0$ we obtain the LMS version (retardation of the first offbeat) and for $\theta_2 < 0$ the SML version (c.f. Fig. 1 (b) and (c)). Then, the model was extended and a third parameter $\theta_3$ was introduced to include the slight variations of the third offbeat. With $\theta_3 = -1$ the pulse lengths are the same as stated in formula (2). But for $\theta_3 = 1$ the second offbeat is on the isochronous time grid. The parameter $\theta_3 = 0$ leads to the same lengths for both offbeats. Binary and ternary swing superimposed leads to the derivation from an isochronous pulse length of

---

[3] With a tempo of $v=120$ BpM (Beats per Minute), the time $t$ in milliseconds for a pulse with a pulse length $p$ given in the unit "normed beat" is computed as $t = 60000p/v$.

$$\mathbf{d} = \frac{1}{6} \begin{pmatrix} \theta_1 + \theta_2 \\ -\theta_1 - \frac{1}{2}(\theta_2\theta_3 + \theta_2) \\ +\theta_1 + \frac{1}{2}(\theta_2\theta_3 - \theta_2) \\ -\theta_1 + \theta_2 \\ +\theta_1 - \frac{1}{2}(\theta_2\theta_3 + \theta_2) \\ -\theta_1 + \frac{1}{2}(\theta_2\theta_3 - \theta_2) \end{pmatrix}. \qquad (3)$$

This formula describes the derivation from the isochronous pulse for a cyclic pattern of six pulses. The vector **1/6** + 1/3**d** describes the normed pulse lengths. [4]

## 2.4 Results

|  | Median of pulse lengths |  |  |  |  |  | $\theta_1$ | $\theta_2$ | $\theta_3$ | $N$ |
|---|---|---|---|---|---|---|---|---|---|---|
| Dundunbé, binary sequence | 0.165 | 0.162 | 0.183 | 0.138 | 0.186 | 0.164 | <u>0.21</u> | <u>-0.26</u> | 0.01 | 62 |
| Dundunbé, ternary sequence | 0.146 | 0.182 | 0.178 | 0.132 | 0.193 | 0.171 | <u>0.10</u> | <u>-0.50</u> | 0.46 | 30 |
| Soli [Polak10] | 0.110 | 0.170 | 0.220 | 0.110 | 0.170 | 0.220 | 0 | -1.02 | -0.88 | - |
| Mendiani [Polak10] | 0.135 | 0.165 | 0.200 | 0.135 | 0.165 | 0.200 | 0 | -0.57 | -1.09 | - |
| Djaa, ternary sequence | 0.194 | 0.141 | 0.168 | 0.181 | 0.159 | 0.158 | 0.13 | 0.37 | 0.62 | 26 |
| Djaa, binary sequence | 0.180 | 0.152 | 0.178 | 0.164 | 0.168 | 0.157 | 0.16 | 0.11 | 0.90 | 32 |
| LMS feel [Polak10] | 0.200 | 0.170 | 0.130 | 0.200 | 0.170 | 0.130 | 0 | 0.62 | -1.08 | - |

Table 1: Measurements of pulse lengths and model parameters: Amount of binary swing $\theta_1$ (values greater than 0 represent offbeat retardation); Amount of ternary swing $\theta_2$ for the first offbeat of a ternary pulse (values greater than 0 represent the LFF feel) and $\theta_3$ for slight variations of the second offbeat. $N$ is the number of measured patterns.

The model parameters $\theta_1$, $\theta_2$ and $\theta_3$ are determined by minimizing the sum of squared errors of the model with respect to the measured pulse lengths. Seed Throwing Optimization [Weede11] was used determined the parameters that minimize the model error. Table 1 shows the median of the measured pulse lengths and the resulting model parameters. In addition, rhythms measured by Polak [Polak10] are included: *Soli* and *Mendiani*, which belong to the SFF feel class like Dundunbé, and Polak's general working hypothesis of 40:34:26 for the LMS feel.

---

[4] The sum of components of vector **d** is one. Thus, the reference normed reference length includes six pulses. Six pulses are three beats in a binary meter, two beats in a ternary meter and 3/2 beats in a quaternary meter. Thus, at a constant tempo the factor 2/3, 3/4 or 2 needs to be multiplied to obtain the real pulse lengths in a given meter.

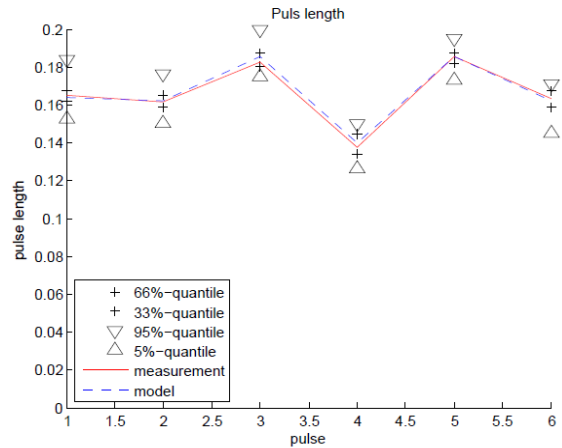

Fig. 2: MMLSLM-Pattern of Dunundbé during a binary sequence.

The 33% and 66% quantiles and the 5% and 95% quantiles are also computed. The predicted pulse lengths of the model are in the 33% and 66% quantiles. This shows that the model is fitting very well to the measurements. Fig. 2 shows the timing profile of a binary sequence of the Dundunbé rhythm. A stable timing pattern of six pulses was observed in all performances. In binary sequences of Dundunbé there is a high degree of binary LS swing superimposed with a ternary SFF feel. In a ternary sequence in the same rhythm, the binary component is decreased to a half amount and the ternary component is increased to a double amount (c.f. underlined values in Table 1). Similar results were observed in Djaa. A binary sequence played in a ternary meter increases the amount of binary swing. However, even a ternary sequence shows, in addition to the ternary swing, a component of binary swing. The superimposed swing pattern of the length six seems to be the norm. Therefore, a measurement of timing patterns should include six pulses, not only two or three.

# 3   Structural analysis of rhythmic patterns

## 3.1   Metric Structure

Fig. 3: Son-Clave and a polymetric interpretation. First row: 4/4-meter; second row: Son-Clave in a sequence of 16 pulses; third row: ternary meter for the first six pulses; fourth row: metric modulation from ternary to binary meter with one ambiguous drum stroke on pulse 6; last row: binary meter for the last 10 pulses.

The user interface of the drum computer allows entering sound events in matrices, like in almost every common sequencer and drum computer. Each matrix in the user interface is associated with one instrument that may contain several timbres. While the rows of the matrix are associated with different timbres of the instrument, the columns represent a time grid made up of 12 or 16 pulses. According to the entered sequence, an appropriate timing pattern should be selected. Therefore, the metric structure is analyzed. For each matrix, a separate analysis is performed.

*Input representation*: The analysis operates on an *event onset vector* of the length 12 or 16 which contains the number of sound events at a pulse. Pulses with no events have a value of zero. An event onset vector is an abstract representation, where beside the order of sound events, no further timing information is included. So, "quantized" input is assumed. For example, the famous Afro-Cuban Son-Clave is described by the event onset vector of (1,0,0,1,0,0,1,0,0,0,1,0,1,0,0,0). Fig. 3 shows the Son-Clave in a simplified music notation. The metric analysis generates a *signature vector* with the same length as the event onset vector. It contains a "2" for a binary sequence and a "3" for a ternary sequence. For example, the signature vector of the Son-Clave could be (3,3,3,3,3,3,2,2,2,2,2,2,2,2,2,2). In this case, the meter of the first six pulses is regarded as ternary, whereas the last 10 pulses are binary.

The aim of the metric analysis is to generate a set of plausible interpretations of the rhythmic sequence. Especially, ambiguities are interesting; plausible interpretations of a rhythmic sequence that have different metric structures. The performance can then be adapted by switching from one interpretation to another. For example, another plausible interpretation of the Son-Clave's meter is (3,3,3,3,3,3,3,3,3,3,3,3,2,2,2,2).

*Event lengths:* The first step in the analysis is to compute an *inter-onset-interval vector* which describes the lengths of the sound events. Its length is the same as the one for the event onset vector. For each sound event, it contains the time difference from a drum stroke to the next. Pulses with no events obtain a value of zero. E. g., the inter-onset-interval vector for the Son-Clave is (3,0,0,3,0,0,4,0,0,0,2,0,4,0,0,0).

*Metric structures:* The second step is to compute metric structures as templates. Temperley [Temperl01] argues that listeners match metric structures with sound events to synchronize with the music and to lay a foundation for the further perceptual organization. He describes the construction of a metric structure by two well-formed rules:

(*Well-formed rule 1 and 2*) *A well-formed metrical structure consists of several levels of beats, such that (1.) Every beat at a given level must be a beat at all lower levels. (2.) Exactly one or two beats at a given level must elapse between each pair of beats at the next level up* (binary vs. ternary subdivision). In his model, a suitable metric structure is aligned with a note sequence and is evaluated by the preference rules. The metric structures serve as templates that are matched with sound events. The most obvious level – the one where most people tap to music, if they are asked for – is called the level of beats. The lowest level of subdivision is called the level of pulses. If a beat is divided into three subdivisions the metric structure is called ternary, if it is divided into two subdivisions the metric structure is called binary.

The metric structure is represented by a *metric vector*. Its components indicate the "beat-strength". For example, one bar of a 4/4 meter, the most common example of a binary metric structure, is represented by the metric vector (5,1,2,1,3,1,2,1,4,1,2,1,3,1,2,1). In this example, the fifth level is the level of the tactus. This metric structure with five levels is visualized in Fig. 3, top row. One bar of a 12/8 meter, the most common example of a ternary meter, is represented by the metric vector (4,1,1,2,1,1,3,1,1,2,1,1). For further computation, all metric vectors are also normed to sum one.

*Local metric structures:* With the preference rules of Temperley the time signature of a piece of music can be computed. Polymetric music can be notated in single time signature like 4/4 or 12/8 meter for the benefit of an easily readable notation. However, in polymetric music each instrument can have its own meter. It can also change from one phrase to another. Here, our aim is to determine these local metric structures. The notated time signature can be called global meter. However, there can be numerous meter changes. To retrieve local metric structures, the preference rules of Temperley are modified. The following preference rules (PR) are suggested and implemented into the drum computer:

PR 1 (*Event Rule*): *Prefer a metric structure that aligns strong beats with sound event-onsets.* This means that a metric structure is preferred that aligns strong beats at points with many drum strokes.

PR 2 (*Length Rule*): *Prefer a metric structure that aligns strong beats with onsets of longer events: large inter-onset-intervals*. This means that a metric structure is preferred which has its strong beats at a point where a drum stroke is followed by a long pause.

For determining local metric structures, a maximum length of one bar (12 or 16 pulses) is taken into account. The minimum length is four pulses because with less than that a binary structure cannot be distinguished from a ternary one. The following metric modulation constraint is introduced: *In a sequence of changing meters, each metric structure is always starting with its strongest beat.* A metric modulation (change of one local meter to another) can occur at any time, however, the following preference rule is introduced:

PR 3 (*Metric modulation*): *Prefer changes in the metric structure (signature boundary) at strong beats of the current structure.* Like harmony changes are preferred to happen on strong beats, so do meter changes. Thus, the next metric structure starts with a strong beat of the previous one. The strong beat is ambiguous because it belongs to both metric structures. In Fig. 3 it is the seventh pulse. The ternary meter is changing into a binary one.

For implementing PR 2 and PR 3, well-formed binary and ternary metric structures of different lengths $n$ are shifted through the cyclic rhythm sequence of the length $L$. For each starting point $p$, ranging from 0 to $L$-1, a score $s(p,n)$ is computed. It describes how much the preference rules are met. For cyclic sequences, the score can be computed as

$$s(p,n) = \sum_{i=0}^{n-1} \left(m_i d_j e_j\right) \text{ with } j = (p+i) \bmod L \;, \tag{4}$$

where $m_i$ is the i-th component of the metric vector of length $n$. $d_j$ is the j-th component of the inter-onset-interval vector of length $L$. $e_j$ is the j-th component of the event-onset vector. The score is computed for local metric structures from length four up to length $L$. For each sequence, the score for a complementing sequence of length $L-n$ is computed and added to the score. For integrating the score of PR 3, the beat strength of the ambiguous pulse at the end of a sequence is determined: the pulse that is already affiliated with the next metric structure. A change on a weak beat is included as a penalty in the score (multiplication with a factor between 0.5 and 1 according to the beat strength). In the example of Fig. 3 it is a beat strength of three for the modulation from ternary to binary and two for the modulation from binary to ternary.

### 3.2 Phrase Structure

According to the Gestalt school we tend to group things together that are similar. In music perception, several similar sound events are grouped into phrases. Here, a simple approach for grouping sound events is chosen:

(*Gap Rule*) According to Tenney et al. [Tenney80] we, *prefer gaps as phrase boundaries (closure), if the gap is larger than the gaps on both sides; especially those ones where the ratio of the gap compared to the previous one is bigger.*

In addition, the following preference rule is stated:

(*Signature rule*) *Prefer to begin phrases where local metric structures are modulated.* This means that there is the tendency to group sound events together that have the same local meter.

## 4 Musical interpretation based on the recognized musical structure

The musical performance of the drum computer is based on the derived metric structure and the derived phrase structure. Therefore, a list of good interpretations (high score according to the preference rules) are regarded. While playing the rhythmic sequence, the currently chosen interpretation is switched from time to time, to create a vivid interpretation. Also, the swing parameters are varied slightly in a smooth continuous way. According to results of the timing measurements presented in this paper, the amount of binary swing is increased for sequences with a local binary meter whereas the amount of ternary swing is decreased. Analog, for ternary sequences the amount of ternary dominates the amount of binary swing. This emphasizes the polymetric phrase and creates tension, an effect of surprise, an intended ambiguity. From the metric vector, an associated velocity pattern is derived, e.g. one that emphasizes strong beats; one that emphasizes the backbeats "2" and "4", or also a pattern that emphasizes the weak beats. In addition to these meter-dependent velocity patterns, the last and the first notes of a phrase are emphasized [Butterf11]. There is also the possibility to annotate an instrument as a solo voice with laid back feel. Butterfield [Butterf11] describes this micro-timing with downbeat retardation in the middle of phrases in detail.

The drum computer is implemented in MAX and Java. Rhythmic patterns can be entered easily in real time. The result sounds natural, lively and vital. It is a very useful tool to learn micro timing patterns.